\documentclass[aps,twocolumn,
showkeys,showpacs,nofootinbib,byrevtex]{revtex4-1}
\usepackage{amsmath,amsfonts,amssymb,amscd,amsxtra,amsthm}
\usepackage{graphicx}
\usepackage{bm}
\usepackage{epstopdf}
\usepackage{epsfig}

\begin{document}

\title{Features of $\omega$ photoproduction off proton target at
backward angles : Role of nucleon Reggeon in $u$-channel with
parton contributions }

\author{Byung-Geel Yu}%
\email{bgyu@kau.ac.kr}%

\author{Kook-Jin Kong}%
\email{kong@kau.ac.kr}%
\affiliation{ Research Institute of Basic Science, Korea Aerospace
University, Goyang, 10540, Korea}


\begin{abstract}
Backward photoproduction of $\omega$ off a proton target is
investigated in a Reggeized model where the $u$-channel nucleon
Reggeon is constructed from the nucleon Born terms in a gauge
invariant way. The  $t$-channel meson exchanges are considered as
a background. While the $N_\alpha$ trajectory of the nucleon
Reggeon reproduces the overall shape of NINA data measured at the
Daresbury Laboratory in the range of $u$-channel momentum transfer
squared $-1.7<u<0.02$ GeV$^2$ and energies at $E_\gamma=2.8$, 3.5
and 4.7 GeV, a possibility of parton contributions is searched for
through the nucleon isoscalar form factor which is parameterized
in terms of parton distributions at the $\omega NN$ vertex.
Detailed analysis is presented for NINA data to understand the
reaction mechanism that could fill up  the deep dip from the
nucleon Reggeon at momentum squared $u=-0.15$ GeV$^2$. The angle
dependence of differential cross sections at the NINA energies
above are reproduced in the overall range of $u$ including the
CLAS data at forward angles in addition. The energy dependence of
the differential cross section is investigated based on the NINA
data and the recent LEPS data. A feature of the present approach
that implicates parton contributions via nucleon form factor is
illustrated in the total and differential cross sections provided
by GRAAL and CB-ELSA Collaborations.
\end{abstract}

\pacs{11.55.Jy, 13.40.-f, 13.60.Le, 12.38.-t}
\maketitle

\section{introduction}

Understanding the structure of hadrons based on QCD is a
longstanding issue and hadron reactions at wide angles are useful
means to investigate parton contributions to reaction processes.
In search of parton distributions in hadrons the recent
experimental activities and theoretical developments have been
concentrated on identifying the scaling of cross sections for
meson photo/electroproductions at mid angle
\cite{brodsky,kroll,goloskokov,dey,zhu} as well as the hadron form
factors in terms of parton densities at large virtual photon
momentum squared \cite{pionff,guidal,diehl}.

In photoproductions of lighter vector mesons $\rho$, $\omega$, and
$\phi$, the peripheral scattering of mesons and Pomeron
exchanges is suppressed at large angles, but an enhancement of
quark exchange is expected due to the smaller impact parameter
$\sim1/\sqrt{-t}$. Scaling of differential cross sections for
hadron reactions with respect to energy is one of the examples
that perturbative QCD predicts at mid angle $\theta\approx
90^\circ$ \cite{brodsky}.
Possibilities of such hard processes were examined in the CLAS
experiment where the differential cross sections of $\omega$
photoproduction were measured over the resonance region
$2.6<W<2.9$ GeV up to a momentum transfer squared $-t=5$ GeV$^2$
\cite{battaglieri}. Around the mid angle $\theta\approx90^\circ$
the scaling of cross sections by $s^8$ in the photon energy
$E_\gamma=3.38-3.56$ GeV seem to be consistent with quark and
gluon exchanges predicted by the QCD-inspired model \cite{laget}
as well as the Reggeized meson exchanges in the $t$-channel with
the trajectory saturated at large $-t$
\cite{bgyu-pi,bgyu-phi,bgyu-delta,bgyu-omega}.

At very backward angles $\theta\approx180^\circ$ beyond
resonances, however, theory and experiment in this kinematical
region are rare \cite{laget,sibirtsev,pire,clifft,battaglieri,morino}
and only the data from
the NINA electron synchroton at the Daresbury Laboratory
\cite{clifft} are available  at present for the $u$-channel
momentum squared $-1.8<u<0.02$ GeV$^2$ at $E_\gamma=2.8$ - 4.7
GeV. Because of the isoscalar nature of $\omega$ meson it is
expected that the reaction at backward angles is dominated by the
$u$-channel nucleon exchange with a dip at $u=-0.15$ GeV$^2$
arising from the nonsense wrong signature zero (NWSZ) of the
$N_\alpha$ trajectory.
Interestingly enough, however, the measured cross section
exhibited the dip much weaker than predicted by the Regge theory,
and hence,
a sort of a mechanism is needed to fill up the depth of the dip.
While there are no other baryon trajectories to play such a role,
the authors of Ref. \cite{clifft} suggested a possibility of
parton contributions there by showing the $s^8$ scaling of cross
section over the dip in the analysis of data at $E_\gamma=3.5$
GeV.

Recently this issue was reexamined in Ref. \cite{sibirtsev} to
investigate backward $\omega$ photoproduction in the baryon pole
model with hadron form factors considered. However, the discussion
on the reaction mechanism around the expected dip as well as its
appearance in the data was no longer valid in the model, because
the occurrence of a dip is a unique feature of the baryon
trajectory at the NWSZ in the Regge theory. Thus, the production
mechanism of $\omega$ photoproduction at very backward angles
remains not fully understood yet, and the topic raised by the NINA
data should be revisited within the Regge framework for the
$u$-channel nucleon exchange.

In this work we study backward $\omega$ photoproduction with our
interest in the search of hard process \cite{laget} involved in
the NINA data \cite{clifft}. Meanwhile, as the scaling of cross
section for meson photoproduction is mostly due to quark and gluon
dynamics through hard process in the midst of meson and baryon
degrees of freedom \cite{kroll,goloskokov}, it becomes, therefore,
an important issue how to consider parton contributions in the
hadronic amplitude of the Regge theory for the present process.

With these in mind, our purpose here is to find a way of considering
parton distributions within the Regge framework for an understanding of
parton contributions to the isotropic cross sections of NINA data
observed in the very backward region.

This paper is organized as follows; Section II devotes to a
construction of  photoproduction amplitude where the nucleon
exchange is reggeized  with the background contribution from the
meson exchanges. Discussion is given on how to consider parton
contributions in the hadronic production amplitude. In Sec. III
numerical consequences in the differential cross sections are
presented to compare with existing data. More proofs for
the validity of the present
approach to current issue are given in the differential and
total cross sections in the region where
the nucleon Reggeon plays a role. Summary and
discussions follow in Sec. IV.

\section{Baryon Reggeon model}

At backward angles where the $u$-channel momentum squared $|u|$ is small
hadron reactions are well described by the
$u$-channel baryon Reggeon in the resonance region \cite{glv}. In
this section we discuss a construction of photoproduction
amplitude for the nucleon Reggeon because the isoscalar $\omega$
prohibits baryon resonances of isospin $I=3/2$ from the $\gamma p\to\omega p$
reaction. In the reggeization of the
relativistic Born terms the nucleon exchange in the $u$-channel
alone is not gauge invariant due to the charge coupling term and
the nucleon exchange in the $s$-channel is introduced further to
preserve  gauge invariance of the production amplitude.

At higher energies beyond the resonance region, the meson exchange
in the $t$-channel begins to give a contribution. To reproduce
experimental data in the overall range of energy and angle,
therefore, it is necessary to include the meson exchange as a
background contribution. However, in order to avoid the
possibility of double counting caused by the $s$ and $t$-channel
duality, if the $t$-channel is reggeized further in addition to
the duality between $s$ and $u$-channel by the $u$-channel
Reggeon, we consider the $t$-channel meson exchanges in the pole
model with the cutoff functions for the divergence of cross
sections at high energies.

\subsection{Nucleon exchange reggeized in the $u$-channel}

\begin{figure}[]
\centering \epsfig{file=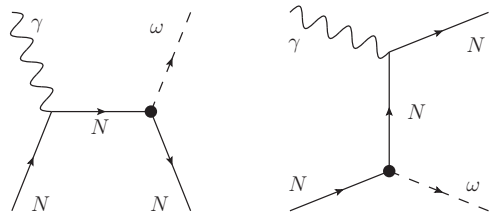, width=0.8\hsize}%
\caption{Nucleon Born terms in $s$- and $u$-channels.
Blobs at the $\omega NN$ vertices
indicate inclusion of the nucleon isoscalar form factor in addition
to the point coupling interaction in Eq. (\ref{17}). } \label{fig1}
\end{figure}

For gauge invariance of nucleon exchange,
as shown in Fig. \ref{fig1}, we now write the nucleon
Born terms as
\begin{eqnarray}\label{born-s}
&&M_s=\bar{u}(p')\Gamma_{\omega NN}{\rlap{/}p+\rlap{/}k+M_N\over
s-M_N^2}\Gamma_{\gamma
NN}u(p),\\
&&M_u=\bar{u}(p')\Gamma_{\gamma
NN}{\rlap{/}p^\prime-\rlap{/}k+M_N\over u-M_N^2}\Gamma_{\omega
NN}u(p),\label{born-u}
\end{eqnarray}
where the electromagnetic and strong coupling vertices are given
by
\begin{eqnarray}\label{vertex}
&&\Gamma_{\gamma NN}=e\left(e_N\rlap{/}\epsilon-{\kappa_N\over
4M_N}\left[\rlap{/}\epsilon,\,\rlap{/}k\right]\right),\label{vertex1}\\
&&\Gamma_{\omega NN}=g_{\omega
NN}\left(\rlap{/}\eta^*+{\kappa_\omega\over
4M_N}\left[\rlap{/}\eta^*,\,\rlap{/}q\right]\right),
\label{vertex2}
\end{eqnarray}
where $u(p)$, $\bar{u}(p')$ are Dirac spinors of the initial and
final nucleons with momenta $p$ and $p'$, and  $\epsilon^\mu$ and
$\eta^{\nu*}$  are polarization vectors of incoming  photon and
outgoing $\omega$ with momenta $k$ and $q$. Charge and anomalous
magnetic moment are $e_N=1$, $\kappa_N=1.79$ for proton and
$e_N=0$, $\kappa_N=-1.91$ for neutron. We take $g_{\omega
NN}=15.6$ by the universality of $\omega$ meson decay constant
$f_\omega$, and $\kappa_\omega=0$ for consistency with the results
from other hadronic process $\gamma N\to\pi^0 N$, for instance
\cite{bgyu-pi}.

The reggeization of the $u$-channel amplitude is simply done by
replacing the $u$-channel pole $(u-M^2)$ with the nucleon Regge propagator
${\cal R}^N(s,u)$, and the reggeized amplitude is expressed as
\begin{eqnarray}\label{nucleon}
&&{\cal M}_N=\left[M_s+M_u\right]\times\left(u-M_N^2\right){\cal
R}^N(s,u),
\end{eqnarray}
with  ${\cal R}^{N}(s,u)$  given by
\begin{eqnarray}\label{regge}
&&{\cal
R}^N(s,u)={\pi\alpha_N'\over\Gamma(\alpha_N(u)+0.5)}{(1+\tau
e^{-i\pi(\alpha_N(u)-0.5)})\over 2\sin\pi(\alpha_N(u)-0.5)}\nonumber\\
&&\hspace{1.5cm}\times\left(s\over s_0\right)^{\alpha_N(u)-0.5},
\end{eqnarray}
where the signature is defined as $\tau=(-)^{J-1/2}$ and $\tau=1$
for nucleon. $s_0=1$ GeV$^2$.

Given the baryon trajectory of the form for the spin $J$
\begin{eqnarray}\label{trajectory}
\alpha(\sqrt{u})=\alpha'u+\alpha_0,
\end{eqnarray}
the MacDowell symmetry predicts the existence of the state with
the same signature $\tau$ but the opposite parity. Thus, denoting
$\alpha^+(\sqrt{u})=\alpha^-(-\sqrt{u})$ to distinguish the states
between the same spin but opposite parities, the nonexistence of
the parity negative state corresponding to the nucleon in the
Chew-Frautschi plot should dictate $1+
e^{-i\pi(\alpha_N(u)(\sqrt{u})-0.5)}=0$ in order not to contribute
to the reaction process, which leads to an occurrence of a dip
with the NWSZ at some position of $u$.

In accordance with many applications we take the slope
$\alpha'=0.9$ GeV$^{-2}$. But the value for the intercept
$\alpha_0$ in literature varies in the range around $-0.3$. Here,
we choose $\alpha_0=-0.365$ so that the trajectory in Eq.
(\ref{trajectory}) yields the dip at $u=-0.15$ GeV$^2$
measured in the NINA data at $E_\gamma=2.8$, 3.5, and $4.7$ GeV
\cite{clifft}.

\subsection{Meson poles in $t$-channel as a background}

Now that the nucleon Reggeon in the $u$-channel in Eq.
(\ref{nucleon}) gives the contribution, in general, by an order of
magnitude smaller than that of the $\pi$ exchange in the
$t$-channel, it is  not enough to reproduce the
cross section data by the nucleon Reggeon contribution alone in the overall
range of angles. Therefore, we
consider the contribution of meson exchanges as a background on which
the nucleon Reggeon is based. As discussed above we treat the
meson exchange simply as the $t$-channel pole with a cutoff
functions and cutoff masses.

For consistency with our previous work, we utilize the meson
exchanges with coupling constants taken the same as in Ref.
\cite{bgyu-omega}. The meson exchanges in Ref. \cite{bgyu-omega}
are now given as the $t$-channel poles with the pole propagator,
\begin{eqnarray}
\left(t-m_\varphi^2\right)^{-1}
\end{eqnarray}
and the cutoff function of the type
\begin{eqnarray}\label{mff}
\left({\Lambda_\varphi^2-m_\varphi^2\over
\Lambda^2_\varphi-t}\right)^n
\end{eqnarray}
for $\varphi=\pi$, $\sigma$, and $f_1$. As for the $f_2$ exchange,
however, due to the highly divergent behavior despite the large
cutoff mass we regard it as the $t$-channel Reggeon  together with
the Pomeron exchange in Ref. \cite{bgyu-omega}. To fix the cutoff
mass $\Lambda_{\varphi}$ in Eq. (\ref{mff}) we exploit the natural
and unnatural parity cross sections over the resonance region.
Before doing this, however, it should be cautioned that the
determination of the sign of the $\pi$ exchange relative to
nucleon  is of importance, because these two are the leading
contributions to the reaction at forward and backward angles,
respectively.

Given the nucleon Reggeon ${\cal M}_N$ in Eq.
(\ref{nucleon}) the production amplitude for the exchanges of the
natural and unnatural parity mesons consists of the following two terms,
\begin{eqnarray}
&&{\cal M}_{\rm nat.}={\cal
M}_N+{\cal M}_\sigma+{\cal M}_{f_2}+{\cal M}_\mathbb{P}\,, \label{main1}\\
&&{\cal M}_{\rm unnat.}={\cal M}_\pi+{\cal
M}_{f_1}\,,\label{main2}
\end{eqnarray}
respectively.

\begin{figure}[h]
\centering \epsfig{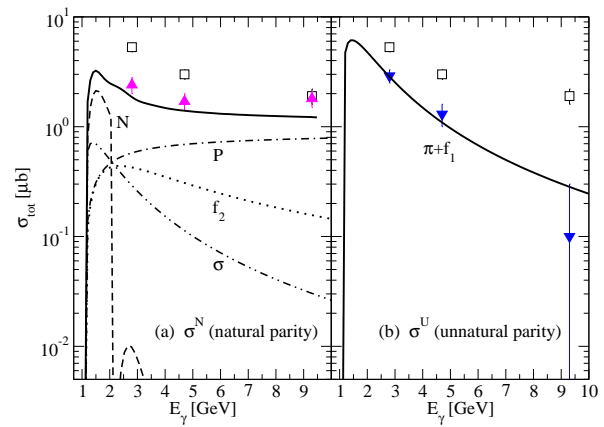}
\caption{Natural and unnatural parity cross sections for $\gamma
p\to \omega p$. Dashed curve is from the proton Reggeon with a
sequence of the dip from the $N_\alpha$ trajectory.
$\kappa_p=1.79$, $g_{\omega NN}=15.6$, $\kappa_\omega=0$ are taken
in Eqs. (\ref{vertex1}) and (\ref{vertex2}). In (b) Solid curve is
from $\pi$ exchange with  $f_1$ of the $10^{-5}$ order
contribution. Data are taken from Ref. \cite{ballam}.}
\label{fig2}
\end{figure}

In Fig. \ref{fig2} (a) by adjusting the cutoff
$\Lambda_\sigma=0.65$ GeV we find that the coupling constant
$g_{\omega NN}=+15.6$  gives a good fit to the natural parity
cross section.
By using $\Lambda_{f_1}=1.3$ GeV for $f_1$ and
$\Lambda_\pi=0.72$ GeV with $n=2$ to suppress the large coupling
at $\gamma\pi\omega$ vertex a fair agreement is obtained as shown in
Fig. \ref{fig2} (b). For further confirmation of
the coupling constants and cutoff masses chosen above
we will provide
differential and total cross sections at low energies
in the following section for numerical consequences. We
summarize the coupling constants and cutoff masses in
Table~\ref{tb1}.

\begin{table}[t]
\caption{Physical constants of exchanged mesons and parameters for
form factors. Masses and cutoff parameters are given in units of
MeV. $f_2$ Reggeon and Pomeron are taken from Ref.
\cite{bgyu-omega}.}
    \begin{tabular}{c|c|c|c|c|cc}
    \hline\hline
        Meson &Mass & $n$ & $\Lambda_\varphi$ & $g_{\gamma \varphi \omega}$ & $g_{\varphi NN}$  \\
        \hline
        $\pi$&134.977& $2$ & $720$ & $-0.69$ & $13.4$ &  \\%
        $\sigma$&500&$1$ & $650$ & $-0.17$ & 14.6 &\\%
        $f_1$ &1281.9 & $1$ & $1300$ & $0.18$ & 2.5& \\%
        \hline
    \end{tabular}\label{tb1}
\end{table}
\vspace{0.5cm}

\subsection{Parton contribution and scaling}

Since our interest is in the analysis of the reaction mechanism
which is suggestive of the hard process at very small $-u$, we
proceed to consider  parton contributions in the hadronic
amplitude in Eqs. (\ref{main1}) and (\ref{main2}).

The differential cross section for the $u$-channel momentum
transfer squared is defined as
\begin{eqnarray}\label{uch-diff}
{d\sigma\over du}
={M_N^2\over16\pi(s-M_N^2)^2}{1\over4}\sum_{\rm spins}\left|{\cal
M}\right|^2.
\end{eqnarray}
The NINA data from the reaction $\gamma p\to\omega p$ at $E_\gamma=3.5$ GeV
are analyzed by a parameterization of $d\sigma/du$ which is
divided by hadronic and hard scattering parts, respectively
\cite{clifft}, i.e.,
\begin{eqnarray}\label{eq13}
{d\sigma\over
du}=\left|A(u)s^{\alpha(u)-1}+B(u)e^{i\phi(u)}s^{-n/2}\right|^2.
\end{eqnarray}
Then, a fit of data with the $s^{8}$ scaling assumed in the hard
scattering term produces the $B(u)$ isotropic at the relative
angle $\phi\simeq 90^\circ$ for all $u$. This suggests incoherency
between hadronic process and hard scattering \cite{clifft}. More
analysis \cite{clifft}  leads us to interpret the hadronic term as
the Reggeon with energy dependence $s^{\alpha-1}$, in which case
the trajectory $\alpha(u)$ generates a typical dip at the expected
position. The $B(u)$ term in the hard scattering is regarded as
parton contributions with $s^8$ scaling with respect to the momentum
squared $u$.

In order to include the parton contributions in the hadronic
amplitude it is natural to suppose a possibility of parton
distribution in nucleon form factors at very backward angles
\cite{guidal} in addition to the point coupling of $\omega NN$
vertex in Eq. (\ref{vertex2}), as depicted in Fig. \ref{fig2}. For
this we introduce the isoscalar form factor $F^{(s)}(u)$ of the
nucleon by the similarity of the $\omega NN$ vertex to a virtual
photon coupling, $\gamma^*NN$. Therefore, the $\omega NN$ vertex
in the nucleon Reggeon in Eq. (\ref{nucleon}) is extended to
include the additional coupling, i.e.,
\begin{eqnarray}\label{17}
&&g_{\omega NN}\gamma^\nu\left({s\over
s_0}\right)^{\alpha_N(u)}\nonumber\\&&\to g_{\omega
NN}\left[\gamma^\nu\left({s\over
s_0}\right)^{\alpha_N(u)}+e^{i\phi(u)}\gamma^\nu
F^{(s)}(u)\left({s\over
s_0}\right)^{\widetilde\alpha_N(u)}\right]\nonumber\\
\end{eqnarray}
with a relative angle $e^{i\phi(u)}$ between the
two coupling phases.
As a result, the nucleon
Reggeon is extended to include parton contributions via the
nucleon isoscalar form factor, ${\cal R}^N\to\left({\cal
R}^N+e^{i\phi(u)}F^{(s)}(u)\widetilde{\cal R}^N\right)$,
and we write the full amplitude as
\begin{eqnarray}\label{full}
{\cal M}=\left({\cal M}_N+{\cal M}_{\rm
b.g.}\right)+e^{i\phi}\widetilde{\cal M}_N,
\end{eqnarray}
in accordance with Eq. (\ref{eq13}) with the two terms in the
first bracket referring to hadronic and the last to parton
contributions, respectively. The term ${\cal M}_{\rm b.g.}$
represents the background coming from all the meson exchanges in
the $t$-channel, i.e., all terms excluding the nucleon term ${\cal
M}_N$ in Eqs. (\ref{main1}) and (\ref{main2}).
In the additional Reggeon $\widetilde{\cal M}_N$ which has the
parton contributions via the form factor, we
further assume that the nucleon trajectory $\alpha_N(u)$
be modified to $\widetilde{\alpha}_N(u)$
in the presence of parton dynamics at very small $|u|$.

The relative angle $e^{i\phi(u)}$ between hadronic and partonic
phases is parameterized as a linear function of $u$
\begin{eqnarray}
\phi(u)=(au+b){\pi \over 180}
\end{eqnarray}
with $a$ in units of GeV$^{-2}$.

The nucleon isoscalar form factor is composed of proton and
neutron charge form factors \cite{perdrisat}
\begin{eqnarray}
F^{(s)}=F_1^p+F_1^n,
\end{eqnarray}
and in the parton model the quark contents of these are expressed
as
\begin{eqnarray}\label{f1ff}
F_1^p=e_u u+e_d d,\ \ F_1^n=e_u d+e_d u
\end{eqnarray}
with the quark distribution \cite{guidal}
\begin{eqnarray}\label{parton}
q(u)=\int dx q_v(x) x^{-(1-x)\alpha'(u-u_0)}
\end{eqnarray}
for the valence quarks $u$ and $d$. Here, we favor to choose the
ansats for the momentum fraction $x$ dependence of partons which
simulates the Regge trajectory with the slope $\alpha'$ in units
of GeV$^{-2}$, because of the relevance to the present formalism.
The momentum squared $u_0$ is the maximum value of $u$ to avoid a
rapid divergence in the region $u > 0$ and $\alpha'$ is considerd to
be a parameter chosen for calculation. As to the valence quark $q_v(x)$
we employ the unpolarized parton distributions for $u$ and $d$
quarks which are
\begin{eqnarray}
&&u_v=0.262x^{-0.69}(1-x)^{3.5}\left(1+3.83x^{0.5}+37.65x\right),\\
&&d_v=0.061x^{-0.65}(1-x)^{4.03}\left(1+49.05x^{0.5}+8.65x\right),\hspace{0.4cm}
\end{eqnarray}
at the input scale $\mu^2=1$ GeV$^2$, respectively. In the
fitting procedure for the relative angle $\phi(u)$ and the
trajectory $\widetilde\alpha_N(u)$ to NINA data we make the slope
parameter adjusted to obtain $\alpha'=0.3$ for a better result.

\begin{figure}[]
\centering
\includegraphics[width=0.9\hsize] {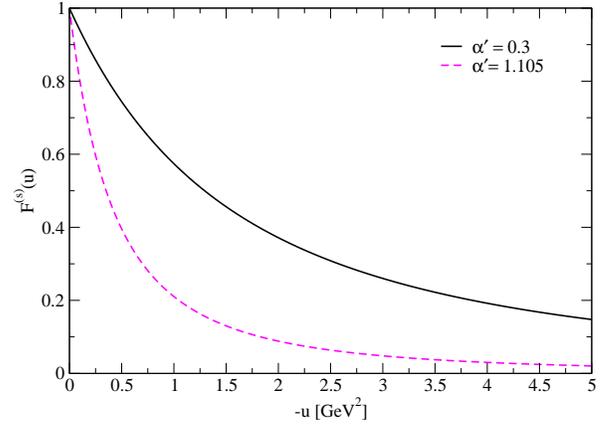}
\caption{$u$ dependence of proton isoscalar form factor
$F^{(s)}(u)$. Solid curve results from $\alpha'=0.3$ GeV$^{-2}$
chosen for the present work and dashed curve from $\alpha'=1.105$
GeV$^{-2}$ used for the fit of proton Dirac form factor
$F_1^p(Q^2)$ in the electroproduction $p(\gamma^*,\, \pi^+n)$
\cite{guidal}.
  } \label{fig3}
\end{figure}

The sensitivity of proton isoscalar form factor $F^{(s)}(u)$ to the
slope parameter $\alpha'$ is examined to discuss its implication
to physical processes. In Fig. \ref{fig3} the form factor with
$\alpha'=0.3$ chosen here is compared to the case with
$\alpha'=1.105$ that is obtained from the fit of proton Dirac form
factor $F_1^p$ to empirical data \cite{tkchoi}. (We mean that the
dependence of the form factor upon the momentum squared $u$ is the
same as the case upon the virtual photon momentum squared $Q^2$.)
Giving the contribution stronger than the case with 1.105, as
shown, the choice of $\alpha'=0.3$  largely deviates from the
empirical form factor with $\alpha'=1.105$ for on-mass shell. In
an application to hadron reactions just as the $\omega$
photoproduction, however, in order to agree with the differential
data at $u=-0.15$ GeV$^2$ the slope $\alpha'=0.3$ is favored
rather than the case of 1.105. This is similar to  the proton
Dirac form factor $F_1^p(Q^2)$ in electroproduction
$\gamma^*p\to\pi^+n$ \cite{tkchoi}, in which case the dipole fit
of the form factor with the cutoff mass $\Lambda=1.55$ GeV is
advantageous to agree with electroproduction data rather than the
form factor with $\alpha'=1.105$. Our choice of $\alpha'=0.3$
yields the form factor which lies between the dipole form factor
above and the one prescribed by Kaskulov and Mosel
\cite{kaskulov}. We regard these results to be feasible, because
the form factor in hadron reactions is, in general, half off-shell
which need not be the same as the on-shell one.

\section{Numerical Results}

\begin{figure}[]
\includegraphics[width=0.9\hsize]{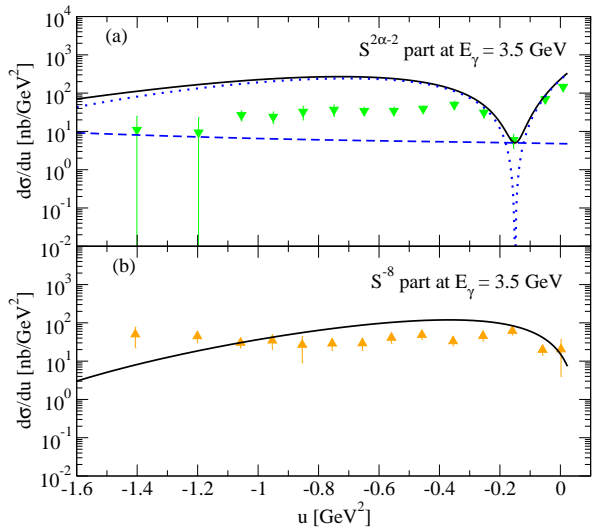}
\caption{Differential cross section for $\gamma p\to\omega p$ at
$E_\gamma=3.5$ GeV in accordance with Eq. (\ref{eq13}). The curve
in (a) results from the hadronic contribution consisting of the
nucleon Reggeon, ${\cal M}_N$ (dotted curve) with the $t$-channel
meson-pole exchange, ${\cal M}_{\rm b.g.}$ (dashed). The curve in
(b) which shows the scaling with respect to momentum squared  $u$ is
resultant of $\widetilde{\cal M}_N$ in Eq. (\ref{full}) with
$\widetilde{\alpha}_N=0.9u-0.56$ and the parton distributions in
the $F^{(s)}(u)$ form factor given in the text. Data are taken from
Ref. \cite{clifft}. } \label{fig4}
\end{figure}

In the fitting procedure to NINA data, our practical purpose is
how to fill up such a deep dip as shown by the dotted curve that
results from the nucleon Reggeon ${\cal M}_N$ in Fig. \ref{fig4}
(a) and how to simulate the scaling behavior of the cross section
with respect to the momentum squared $u$ in (b) with
parton contributions from the nucleon  Reggeon $\widetilde{\cal
M}_N$ in addition. However, there is no evidence for a dip in the
cross section of (b) for which the additional Reggeon
$\widetilde{\cal M}_N$ is applied, while it should produce a dip
at the same place of $u$ by the NWSZ of the trajectory
$\widetilde{\alpha}_N(u)$, unless different from the
$\alpha_N(u)$. Thus, expecting not only the scaling without a dip
in the hard scattering but also the reaction mechanism to fill up
the depth of the original dip by a destructive interference between
hadronic and hard processes, we
have to move the position of the dip by $\widetilde{\cal M}_N$ to
another place by altering the trajectory $\widetilde{\alpha}_N$ in
Eq. (\ref{17}). Hence, we let the trajectory of the
$\widetilde{\cal R}^N$ vary in the fitting procedure to obtain
\begin{eqnarray} \widetilde\alpha_N(u)=0.9\,u-0.56
\end{eqnarray}
with the intercept adjusted for the best fit to the NINA data in
the overall range of energy and angle. The dip position of
$\widetilde{\cal R}^{N}$ is now at $u=+0.067$ GeV$^2$, and
hence, not appearing
in the kinematical region of the reaction
$u<0.02$ GeV$^2$ so that the resulting cross section in (b)
could simulate the scaling without a dip, as shown by the solid
curve.

Given the differential cross section $d\sigma/du$ for $\gamma p\to
\omega p$ at $E_\gamma=3.5$ GeV separately into two parts as in
Eq. (\ref{eq13}), the hadronic contribution corresponding to the
$s^{\alpha(u)-1}$ term and the $s^{-8}$ scaling by parton
contributions are shown in Fig. \ref{fig4} (a) and (b),
respectively. In (a) the solid curve results from the sum of the
dotted curve by the nucleon Reggeon and the dashed one by the
$t$-channel meson exchanges, i.e., from the hadronic amplitude
${\cal M}_N+{\cal M}_{\rm b.g.}$ in Eq. (\ref{full}).  As
expected, the Reggeon ${\cal M}_N$ produces the deep dip at
$u=-0.15$ GeV$^2$ some part of which could be covered over
with the meson contribution. Nevertheless, the dip remains not
fully compensated yet. Moreover, the solid curve of the hadronic
contributions ${\cal M}_N+{\cal M}_{\rm b.g.}$ is overestimating
the data over $0.3$ GeV$^2$ $<|u|$, which should be reduced.
In (b) the solid curve is from the additional Reggeon,
$\widetilde{\cal M}_{N}$ with the trajectory
$\widetilde\alpha_N(u)=0.9\,u-0.56$ with the dip at $u=+0.067$
GeV$^2$ so that the $\widetilde{\cal M}_N$ could preserve the
cross section data without dip near $u\approx0$. On the
other hand, this term, when combined with the hadronic part ${\cal
M}_N+{\cal M}_{\rm b.g.}$ via the relative angle
$\phi\simeq90^\circ$, should fill up the rest of the part of the dip as
well as it should reduce the overestimating contribution of the
solid curve to hadronic data in (a).

\begin{figure}[]
\includegraphics[width=0.9\hsize]{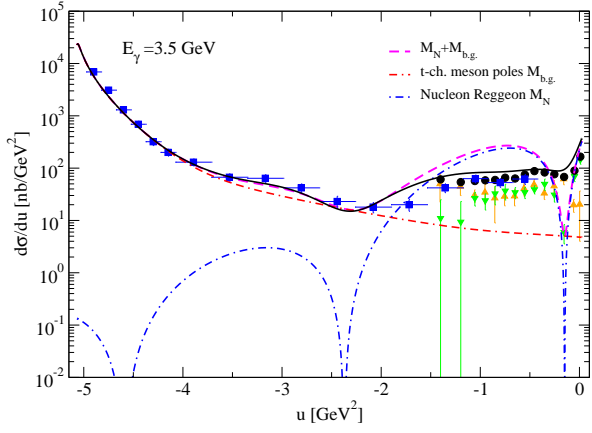}
\caption{Differential cross section for $\gamma p\to\omega p$ at
$E_\gamma=3.5$ GeV.  The dashed curve corresponds to the solid one
in Fig. \ref{fig4} (a). The solid curve from the full amplitude
shows a good agreement with data in the overall range of $u$ with
the fit of $a=-65$, $b=93$ for $\phi(u)$.  Data are collected from
Refs. \cite{battaglieri} and \cite{clifft}.  } \label{fig5}
\end{figure}
\begin{figure}[]
\centering \epsfig{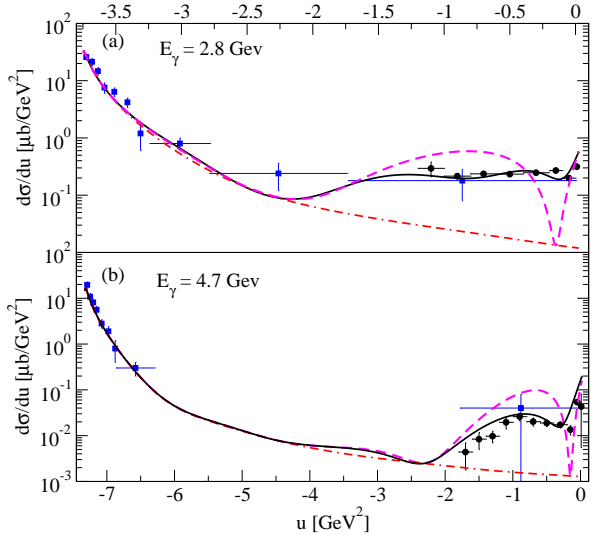}
\caption{Differential cross sections $d\sigma/du$ at
$E_\gamma=2.8$ and $4.7$ GeV with the parton contributions in Eq.
(\ref{full}) included. The cross section in the upper panel is
fitted with $a=-95$, $b=70$ and lower with $a=-35$, $b=120$ for
$\phi(u)$.  Notations are the same as in Fig. \ref{fig5}. Data are
taken from Refs. \cite{ballam,clifft}. } \label{fig6}
\end{figure}

Figure \ref{fig5} shows the differential cross section
$d\sigma/du$ for $\gamma p\to\omega p$ when the cross sections (a)
and (b) of Fig. \ref{fig4}  are combined with each other. The
parameters $a=-65$ and $b=93$ are chosen for the relative angle
$\phi(u)$ between hadronic and parton phase of the reaction
process in Eq. (\ref{eq13}). In actual, these parameters
correspond to $\phi=0.57\pi$ at $u=-0.15$ GeV$^2$ which is
close to $\phi\simeq 90^\circ$ from the decomposition of the NINA
data in Fig. \ref{fig3}. The additional Reggeon $\widetilde{\cal
M}_N$ with parton contributions  vanishes over $|u|> 2$
GeV$^2$. The respective roles of the Reggeon ${\cal M}_N$ and
the meson exchanges ${\cal M}_{\rm b.g.}$ in the $t$-channel are
depicted by the dash-dotted and dash-dash-dotted curves in order.
The rapid increase of cross sections in the NINA data at large
$|u|$ are identified by the $t$-channel meson exchanges, while the
Reggeon ${\cal M}_N$ gives the contribution in the backward region
$|u|<2.5$ GeV$^2$, as expected. We further note that these
hadronic contributions, ${\cal M}_N+{\cal M}_{\rm b.g.}$ reproduce
the shape convex up in the interval $-3.5<u<-2.5$ GeV$^2$
which was once described by the two quarks exchange in previous work as
elaborated in Ref. \cite{laget}. In the current approach since the
dip position of $\widetilde{\cal R}^N$ is removed out to the kinematical
region, the term $\widetilde{\cal M}_N$
could play the role to fill up some part of the original dip at
$u=-0.15$ GeV$^2$ as well as to reduce the overestimating
contribution of the hadronic part (dashed curve) in Figs. \ref{fig5}
and \ref{fig6} up to $ u\approx -2$ GeV$^2$, when combined
with each other through $\phi\simeq90^\circ$.

Differential cross sections $d\sigma/du$ at $E_\gamma=2.8$ and
$4.7$ GeV are presented in Fig. {\ref{fig6}} to show a fair
agreement with data. The cross section at $E_\gamma=2.8$ GeV in
the upper panel is fitted with $a=-95$ and $b=70$  and the cross
section in the lower panel with $a=-35$ and $b=120$ for $\phi$,
respectively. These parameters at $u=-0.15$ GeV$^2$, for
instance, yield $\phi\approx0.47\pi$ and $\phi\approx0.7\pi$, in
order. Thus, the relative angle $\phi$ is not quite unique to
energy dependence of cross sections and we have to change the
parameters $a$ and $b$ to accomplish such an agreement with data
at the photon energies $E_\gamma=2.8$, 3.5, and 4.7 GeV given. The
contribution of meson exchanges as shown by the dash-dash-dotted
curve is responsible for the forward enhancement of the cross
section at large $|u|$, as before. Solid curve shows the
improvement of hadronic contribution (dashed curve) by the
additional Reggeon $\widetilde{\cal M}_N$ with parton
contributions at backward angles.

\begin{figure}[]   
\centering
\includegraphics[width=1.0\hsize]{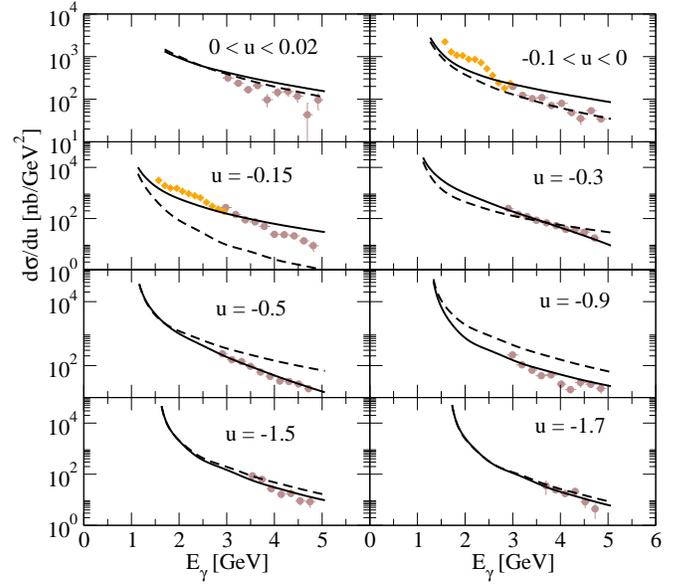}
\caption{Energy dependence of differential cross sections
$d\sigma/du$. The solid curve is reproduced by a linear fit of
data with $a=31{\rm GeV}^{-1}E_\gamma-178$ and
$b=26{\rm GeV}^{-1}E_\gamma-0.64$ for the
relative angle $\phi$. The dashed curve is without parton
contributions. The momentum squared $u$ is given in units of
GeV$^2$. Data over $E_\gamma\approx 3$ GeV are taken from Ref.
\cite{clifft} and at low energies are taken from Ref.
\cite{morino}.} \label{fig7}
\end{figure}

To test the validity of the present approach further  we check up
the energy dependence of differential cross sections in the
range  $-1.8<u<0.02$ GeV$^2$ and present the result in
Fig. \ref{fig7}. The NINA cross sections $d\sigma/du$ in eight angle
bins are reproduced up to photon energy $E_\gamma=5$ GeV.
In the differential cross sections at $-0.1 < u < 0$ and $u=-0.15$
GeV$^2$ the data recently measured at the SPring-8/LEPS facility
\cite{morino} are included further for relevance to the present analysis.
Since the angle $\phi$ has the energy dependence as shown in Figs.
\ref{fig5} and \ref{fig6}, we make parameterized $a=31{\rm GeV}^{-1}
E_\gamma-178.2$ and $b=26{\rm GeV}^{-1}E_\gamma-0.64$ for $\phi$ as a linear
function of photon energy. It is interesting to see the peculiar
behavior of the cross section at $u=-0.15$ GeV$^2$.
The difference between the contributions of nucleon Reggeon
with and without partons via the form factor $F^{(s)}$
becomes maximal around the dip position at $u=-0.15$ GeV$^2$ and
vanishes over $u\approx-1.7$ GeV$^2$, as expected.
With an expectation that the exponent $x$ of $As^{-x}$ term
converted to an effective trajectory $\alpha(u)$, as indicated in
Ref. \cite{clifft},  the role of new trajectory
$\widetilde\alpha_N(u)$ with parton contributions is significant
at $u=-0.15$ GeV$^2$, although we do not follow the negative power
of $x$ assumed there.
The analysis of Fig. \ref{fig7} is
consistent with the observation of Ref. \cite{clifft} that the
slope of the cross section at $u=-0.15$ GeV$^2$ is different from
others.
We obtain a reasonable result in cross sections that cover the
overall range of energy and momentum of NINA data. These findings
further support the validity of the present approach.

\begin{figure}[]
\centering \epsfig{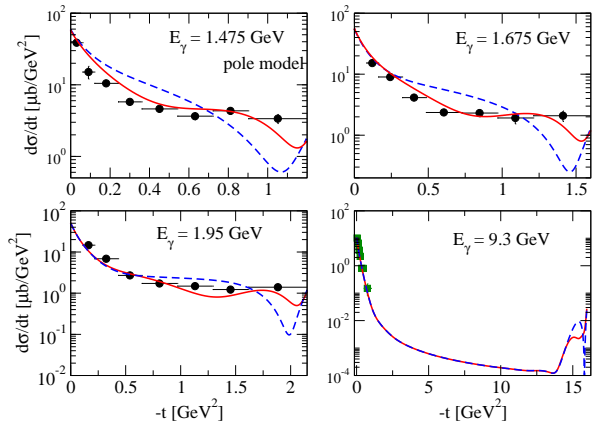}
\caption{Differential cross sections for $\gamma p\to \omega p$ in
low energy region below $E_\gamma=2$ GeV and at high energy
$E_\gamma=9.3$ GeV. Solid curves are from the full calculation
with parton contributions. Dashed curves are without parton
contributions. Data below 2 GeV are taken from Ref. \cite{dietz}
and data at 9.3 GeV from Ref. \cite{ballam}.
  } \label{fig8}
\end{figure}

\begin{figure}[]
\centering%
\epsfig{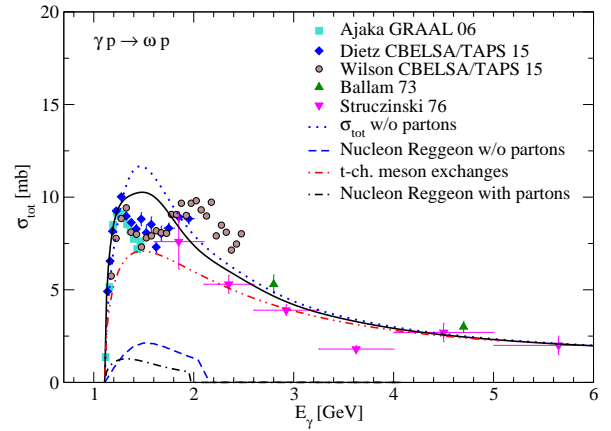}%
\caption{Total cross section for $\gamma p\to \omega p$. Cross
sections $\sigma_{tot}$ with and without parton contributions are
given by solid and dotted curves in response to the nucleon
contributions with and without isoscalar form factor which are
given by dashed and dash-dotted curves. The background
contribution  from the $t$-channel exchange is depicted by
dash-dot-dotted curve. Data are taken from Refs.
\cite{ballam,dietz,wilson,ajaka,struczinski}. } \label{fig9}
\end{figure}

Predictions for differential and total cross sections are
presented in Figs. \ref{fig8} and \ref{fig9} with the
parameters  $a$ and $b$ in Fig. \ref{fig7} for the
angle $\phi$. The differential cross sections are reproduced in
the overall range of $-t$ at the given energy $E_\gamma$. In Fig.
\ref{fig8} each cross section at very backward angles reveals the
role of the nucleon isoscalar form factor with partons
to make the model prediction closer to
experimental data even in the low energy region less than
$E_\gamma=2$ GeV. Within the present framework it is rather
natural that parton contributions via the form factor
could appear at lower energies
where the nucleon Reggeon is dominant. Such an evidence of parton
distributions at low energies can be traced out in the total cross
section as presented up to $E_\gamma=6$ GeV in Fig. \ref{fig9}.
The nucleon isoscalar form factor with
parton distributions plays the role to reduce the cross section.
Thus, it is a feature of the present approach that the parton
distribution could play a role through hadron form factors without
deep scattering of virtuality photon at high energies.
Together with nucleon Reggeon, a fair
agreement with data on total and differential cross sections
further confirms the validity of the background contribution with
cutoff masses chosen for the present calculation.


\section{Summary and discussions}

Backward photoproduction of $\omega$ meson off a proton target is
investigated within the Regge framework where the nucleon Born
terms in the $s$- and $u$-channels  are reggeized for the gauge
invariant $u$-channel nucleon Reggeon.  The exchanges
$\sigma+\pi+f_1+f_2+Pomeron$ in the $t$-channel are included as a
background contribution to reproduce reaction cross sections. The
cutoff masses for the cutoff functions for  $\sigma+\pi+f_1$ poles
in the $t$-channel are determined from natural and unnatural parity cross
sections  as shown in Fig. \ref{fig2}.

While the $N_\alpha$ trajectory of the nucleon Reggeon reproduces
the overall shape of the NINA data measured at the Daresbury
Laboratory in the range of $-1.7<u<0.02$ GeV$^2$ and
energies at $E_\gamma=2.8$, 3.5 and 4.7 GeV, a possibility of
parton contributions is searched for by considering the nucleon
isoscalar form factor at the $\omega NN$ vertex which is
parameterized in terms of parton distributions. The nucleon
Reggeon would make a deep dip at $u=-0.15$ GeV$^2$, which
should be covered over by a fill-up mechanism in order to agree
with the cross sections observed in the Daresbury experiment. The
dip of the nucleon Reggeon is covered over partly with the meson
exchanges and partly with the additional Reggeon with partons at
very small $|u|$. From the practical point of view a manipulation
of the relative phase $\phi\simeq90^\circ$ and the need of the
$t$-channel contributions are of importance to reproduce the NINA
data. Due to the parton densities in the nucleon isoscalar form
factor, the parton contributions through the nucleon Reggeon
activates rather in the lower energy region than is usually
expected, as shown by the description of total and differential
cross sections of GRAAL and CB-ELSA Collaborations.

Confronting the impending 12 GeV upgrade of CLAS detector, it is
timely interesting to observe in experiments how partons would
manifest  themselves in the midst of hadronic degrees of freedom
and how we could understand such a phenomenology through
theoretical analysis. In this work we have illustrated how to
incorporate parton distributions with hadronic degrees of
freedom in hadron models just as the Regge model and the
prescription we favored here offers an intuitive way to consider
parton contributions in hadron reactions via the hadron form
factors. Together with the scaling of meson photoproduction by the
factor $s^7$ around mid angle $\theta=90^\circ$ observed in the
Jefferson Lab, the NINA data at Daresbury, though almost a 30
years-old issue, provide information further for our understanding
of quark dynamics inside hadrons  in the limit $-u\approx 0$ that
we could expect to observe in future experiments at the 12 GeV
upgraded CLAS as well as those facilities LEPS, CB-ELSA, and GRAAL.

       \section*{Acknowledgments}
This work was supported by the National Research Foundation of
Korea Grant No. NRF-2017R1A2B4010117.
%
\\

\end{document}